\documentclass[prb, preprint,unsortedaddress,showpacs]{revtex4}

\usepackage{amsfonts, amsbsy, amssymb, amsmath, graphicx,  float} 

\usepackage{color}
\usepackage{float}
\usepackage{amsthm}
\restylefloat{figure}
\usepackage{pifont}      
\reversemarginpar

\begin{document}


\title{Phase space barriers and dividing surfaces in the absence of  
critical points of the potential energy: Application to roaming in ozone}



\author{Fr\'ed\'eric A. L. Maugui\`ere}
\email[]{frederic.mauguiere@bristol.ac.uk}
\affiliation{School of Mathematics \\
University of Bristol\\Bristol BS8 1TW\\United Kingdom}

\author{Peter Collins}
\email{peter.collins@bristol.ac.uk}
\affiliation{School of Mathematics \\
University of Bristol\\Bristol BS8 1TW\\United Kingdom}

\author{Zeb C. Kramer}
\email[]{zebcolterkramer@gmail.com}
\affiliation{Department of Chemistry and Chemical Biology\\
Baker Laboratory\\
Cornell University\\
Ithaca, NY 14853\\USA}

\author{Barry K. Carpenter}
\email{carpenterb1@cardiff.ac.uk}
\affiliation{School of Chemistry \\
Cardiff University\\Cardiff CF10 3AT\\United Kingdom}

\author{Gregory S. Ezra}
\email[]{gse1@cornell.edu}
\affiliation{Department of Chemistry and Chemical Biology\\
Baker Laboratory\\
Cornell University\\
Ithaca, NY 14853\\USA}

\author{Stavros C. Farantos}
\email{farantos@iesl.forth.gr}
\affiliation{Institute of Electronic Structure and Laser, Foundation for Research and Technology - Hellas, and \\
Department of Chemistry, University of Crete, Iraklion 711 10, Crete, Greece }

\author{Stephen Wiggins}
\email[]{stephen.wiggins@mac.com}
\affiliation{School of Mathematics \\
University of Bristol\\Bristol BS8 1TW\\United Kingdom}


\date{\today}

\begin{abstract}

We examine the phase space structures that govern reaction dynamics 
in the absence of critical points
on the  potential energy surface. We show that in the vicinity of hyperbolic invariant tori 
it is possible to define phase space dividing surfaces  
that are analogous to the dividing surfaces governing 
transition from reactants to products near a critical point of the  potential energy surface.
We investigate the problem of capture of an atom by a diatomic molecule 
and show that a normally hyperbolic invariant manifold exists at large
atom-diatom distances, away from any critical points on the
potential. This normally hyperbolic invariant manifold is the anchor for the construction of a 
dividing surface in phase space,
which defines the outer or loose transition state governing capture dynamics.
We present an algorithm 
for sampling an approximate capture dividing surface, and
apply our methods 
to the recombination of the ozone molecule. We treat
both 2 and 3 degree of freedom models with zero total angular momentum. 
We have located the normally hyperbolic invariant manifold  from which the  orbiting (outer) transition state is constructed. This forms the basis for our  analysis of  trajectories for ozone in general, but with particular emphasis on the roaming trajectories.

\end{abstract}



\maketitle


\section{Introduction}
\label{sec:intro}

Critical points of the potential energy surface (PES) have played, and continue to play, 
a significant role in how one thinks about transformations of physical systems 
\cite{Mezey87,Wales03}. The term `transformation' may refer to chemical 
reactions such as isomerizations 
\cite{Chandler78,Berne82,Davis86,Gray87,Minyaev91,Minyaev94a,Baer96,Leitner99,Waalkens04,Carpenter05,Joyeux05,Ezra09a}
or the analogue of phase transitions for finite size systems \cite{Wales03,Pettini07,Kastner08}.
A comprehensive description 
of this so-called `energy landscape paradigm' is given in Ref.\ \onlinecite{Wales03}. 
The energy landscape approach is an attempt to understand  dynamics in the context of the geometrical 
features of the  potential energy surface, i.e., a configuration space approach. 
However,  the arena for dynamics is phase space \cite{Arnold06,Wiggins_book03,Farantos_2014Springer}, and numerous  
studies of nonlinear dynamical systems have 
taught us that  the rich variety of dynamical behavior possible in nonlinear systems {\em cannot} 
be inferred from geometrical properties of the potential energy surface alone;
An instructive example is the fact that the well-studied and nonintegrable
H\'enon-Heiles potential can be obtained by series expansion of the 
completely integrable Toda system \cite{Lichtenberg92}.
Nevertheless, the configuration space based landscape paradigm is physically 
very compelling, and there has been a great deal of recent work describing 
{\em phase space signatures} of index one saddles \cite{saddle_footnote1}
of the potential energy surface that are relevant to reaction dynamics
(see, for example, Refs 
\onlinecite{Komatsuzaki00,wiggins2001impenetrable,uzer2002geometry,waalkens2007wigner}).
More recently, index two \cite{Heidrich86,Ezra09,haller10} and higher index \cite{Shida05} saddles
have been studied (see also Refs \onlinecite{ezra2009phase,collins:244105,Mauguiere13}).

The work on index one saddles has shown that, in {\em phase space}, the role of the saddle {\em point} 
is played by an {\em invariant manifold} of saddle stability type, 
a so-called normally hyperbolic invariant manifold or NHIM \cite{Wiggins90,Wiggins94}. 
The NHIM proves to be the anchor 
for the  construction of dividing surfaces (DS) that have the  properties of  no (local) recrossing 
of trajectories and minimal (directional) flux \cite{waalkens2004direct}.
There is an even richer variety of phase space structures and invariant manifolds associated 
with index two saddles of the potential energy surface, and their implications for 
reaction dynamics have begun to be explored \cite{Ezra09,collins:244105,Mauguiere13}. 
Fundamental theorems assure the existence of
these phase space structures and invariant manifolds for a range of energy above 
that of the  saddle \cite{Wiggins94}.  
However, the precise extent of this range, as well as the nature 
and consequences 
of any bifurcations of the phase space structures and invariant manifolds 
that might occur as energy is increased, is not known and is a topic of continuing research
\cite{li2009bifurcation,Inarrea11,mauguiere2013bifurcations,mackay2014bifurcations}. 

While work  relating phase space structures and invariant manifolds to saddle 
points on the potential energy surface has provided new insights and techniques for 
studying reaction dynamics \cite{Komatsuzaki00,wiggins2001impenetrable,uzer2002geometry,waalkens2007wigner},  
it by no means  exhausts all of the  rich possibilities of 
dynamical phenomena associated with reactions. 
In fact, a body of work has called into question the 
utility of concepts such as the reaction path and/or 
transition state (TS) 
\cite{Sun02,townsend2004roaming,Bowman06,Lopez07,Heazlewood08,Shepler08,suits2008,bowman2011roaming,Bowman2011Suits,BowmanRoaming}.
Of particular interest for the present work is the recognition that there are 
important classes of chemical reaction,
such as ion-molecule reactions and association reactions in barrierless systems, for which
the transition state is not necessarily
directly associated with the presence of a saddle point on the potential energy surface;
Such transition states might be generated dynamically,
and so are associated with critical points of the amended or effective
potential, which includes centrifugal contributions to the energy 
\cite{Wiesenfeld03,Wiesenfeld04a,Wiesenfeld05}.
The phenomenon of transition state switching in ion-molecule reactions 
\cite{Chesnavich1981,Chesnavich82,Chesnavich1986}
provides a good example of the dynamical complexity possible in such systems,
as does the widely studied phenomenon of roaming \cite{townsend2004roaming,suits2008,BowmanRoaming}.

The lack of an appropriate critical point on the potential 
energy surface with which to associate a dividing surface separating
reactants from products in such systems does {\em not} however mean that there are no relevant
geometric structures and invariant manifolds in {\em phase space}. 
In this paper we  discuss the existence of NHIMs, 
along with their stable and unstable manifolds and associated dividing surfaces, 
in regions of phase space that do not correspond to saddle points of the potential energy surface. 
We describe
a theoretical framework for describing and computing such NHIMs.
Like the methods associated with index one and two saddles, the method we develop for 
realizing the existence of NHIMs is based on normal form theory; however, rather than normal 
form theory for  saddle-type equilibrium points of Hamilton's equations 
(which are the phase space manifestation of index one and two saddles of the potential energy surface), 
we use normal form theory for  certain hyperbolic invariant tori.  
The  hyperbolic invariant tori (and their stable and unstable manifolds) alone are not adequate, 
in terms of their dimension, for constructing NHIMs that have 
codimension one stable and unstable manifolds (in a fixed energy surface). 
However, by analogy with the use of index one  saddles 
to infer the existence of NHIMs (together with their stable and unstable manifolds, 
and  other  dividing surfaces having appropriate dimensions),  these particular hyperbolic
invariant tori can likewise be used to  infer the existence of phase space structures 
that \emph{are} appropriate for describing  reaction dynamics in situations where there is no 
critical point of the potential energy surface in the relevant region of configuration space. 

This paper is organized as follows:
Section \ref{sec2} presents the general theoretical framework for
describing phase space structure in the vicinity of partially hyperbolic tori.
We describe the associated NHIM and corresponding (codimension one) DS on the energy shell.
In Sec.\ \ref{sec3}, we apply these ideas to a model for atom-diatom collision dynamics;
specifically, capture of an atom by passage through an outer or `loose' 
transition state defined by the general theory of Sec.\ \ref{sec2}.
By making a number of simplifying assumptions, we are able to formulate 
a practical algorithm for sampling the relevant phase space DS so that
trajectory initial conditions on the DS can be obtained in a systematic way.
Section\ \ref{sec4} applies this algorithm to sample the DS in a model for the
initial recombination reaction in ozone, O $+$ O$_{2}$ $\rightarrow$ O$_{3}$.
Both 2 and 3 DoF models are considered.  An interesting finding is that
roaming dynamics (as defined and explored in our earlier work
\cite{Mauguiere_2014cpl,Mauguiere_2014jcp,Mauguiere_2014tca,Mauguiere_2015jpcl}) 
is important in this reaction, and may therefore be of significance in 
understanding the anomalous isotope effect 
\cite{janssen2001kinetic,gao2001strange,gao2002theory,gao2002theoretical,schinke2006dynamical}.
Section \ref{sec:conclusions} concludes.

\newpage

\section{Reaction dynamics in the absence of equilibrium points}
\label{sec2}

Here we discuss the work of  Ref.\ \onlinecite{bolotin2000remarks}
on partially hyperbolic tori in Hamiltonian  systems and associated phase space structures. 
The starting set-up is an $m$ DoF canonical Hamiltonian system defined on 
a $2m$-dimensional phase space, which we  can take to be $\mathbb{R}^{2m}$.

The steps involved in describing phase space structure in the vicinity of
partially hyperbolic tori are as follows:

\medskip
\noindent
{\bf Step 0: Locate a partially hyperbolic invariant torus in the original system.} 
First we define the term ``partially hyperbolic invariant torus''.
A  hyperbolic periodic orbit provides the simplest example.
The mathematics literature contains rather technical 
discussions of the notion of partial hyperbolicity \cite{bolotin2000remarks}, 
but a practical definition  is that, under the linearized dynamics, 
in directions transverse to the torus {\em and} the directions 
corresponding to the variables canonically conjugate to the angles 
defining the torus (this is where the qualifier ``partially'' comes from),  
there is an equal number of exponentially growing and exponentially decaying directions.
The following steps express 
the original Hamiltonian in coordinates defined near the torus. 

\medskip
\noindent
{\bf Step 1: Express the Hamiltonian in coordinates near the  
partially hyperbolic invariant torus.} 
\medskip

Following Ref.\ \onlinecite{bolotin2000remarks}  we denote the partially 
hyperbolic $n$-dimensional invariant torus by $N$.
In Theorem 5 of Ref.\ \onlinecite{bolotin2000remarks} 
(with the explicit construction of coordinates in Sec. 5), 
Bolotin and Treschev show that there exist canonical coordinates 
in a neighborhood of $N$ in which the Hamiltonian takes the following form:
\begin{equation}
H(\theta, I, z_+, z_-) = \langle \omega, I \rangle + \frac{1}{2} \langle AI, I \rangle + \lambda (\theta)  z_-  z_+  + {\cal O}(3) , \quad \theta \in \mathbb{T}^n, \, I \in \mathbb{R}^n, \, z_- \in \mathbb{R}^1, \, z_+ \in \mathbb{R}^1,
\label{BTham}
\end{equation}
where $\langle \cdot , \cdot \rangle$ denotes the usual Euclidean 
inner product on $\mathbb{R}^n$, $A$ denotes an $n \times n$ symmetric matrix, 
and $\lambda (\theta) > 0$ for all $\theta \in \mathbb{T}^n$ 
(this is the condition insuring hyperbolicity).
The  corresponding Hamiltonian vector field is
\begin{subequations}
\label{BThameq}
\begin{align}
\dot{\theta} & =  \omega + AI + {\cal O}(2), \\
\dot{I} & =  {\cal O}(2), \\
\dot{z}_+ & =  \lambda (\theta) z_+ + {\cal O}(2),  \\
\dot{z}_- & =  -\lambda (\theta)  z_-  + {\cal O}(2).
\end{align}
\end{subequations}
Note that in these coordinates the invariant torus is given by:
\begin{equation}
N = \left\{ (I, \theta, z_+, z_-) \, \vert \, I=z_+ = z_- =0 \right\}
\end{equation}

Several comments are now in order.

\begin{itemize}

\item It is important to realize that \eqref{BTham} is not  the ``normal form'' 
that is usually discussed in Hamiltonian mechanics \cite{Meyer2009}. 
This is because $\lambda (\theta)$ is not necessarily constant. 
A normal form requires additional transformations to 
transform $\lambda (\theta)$ to a constant (the Floquet transformation). 
We will see that such a transformation
is not necessary for constructing NHIMs and DSs.

\item From the form of the equations \eqref{BThameq} the origin of the term 
``partially'' hyperbolic is clear: not all directions transverse 
to the torus are hyperbolic (the transverse directions that are not hyperbolic  
are those described by the coordinates $I$). 

\item For $n=1$, $N$ is a periodic orbit.

\item For $n>1$ the frequency vector of $N$, $\omega$, must be 
\emph{nonresonant}. More precisely, it is required to satisfy a 
diophantine condition \cite{Wiggins_book03}, i.e. 
\begin{equation}
\vert \langle \omega , k \rangle \vert > \alpha \vert k \vert^{-\beta}, \quad \alpha, \, \beta >0, \, k \in \mathbb{Z}^n -\{0 \},
\end{equation}
where $\vert k \vert \equiv \sum_{i=1}^n \vert k_i \vert$.

\item $z_-$ and $z_+$ are the exponentially  decaying and growing directions, 
respectively, normal to $N$; They are of equal dimension.  
Here we are only considering the case where they are 1-dimensional. 
This is the analogous case to index one saddles for equilibria. 
Both $z_{-}$ and $z_{+}$ can each have dimension greater that one.  
In that case we would  have a generalization of the notion of index $k$ saddles \cite{collins:244105}, 
for $k \ge 1$, to periodic orbits and tori. We will not consider that case here.

\end{itemize}

It is useful to emphasize at this point that the phase space is $2n + 2 \equiv 2m$ dimensional.

Before using the Hamiltonian  \eqref{BTham} to construct  NHIMs and DSs we  
(symplectically) rotate the  saddle coordinates in the usual way:
\begin{equation}
\label{eq:saddlerot}
z_+ = \frac{1}{\sqrt{2}} (p-q), \quad z_- = \frac{1}{\sqrt{2}} (p+q),
\end{equation}
to obtain the Hamiltonian:
\begin{equation}
H(\theta, I, q, p) = \langle \omega, I \rangle + \frac{1}{2} \langle AI, I \rangle + \frac{\lambda (\theta)}{2}  (p^2 - q^2 ) +{\cal O}(3), \quad \theta \in \mathbb{T}^n, \, I \in \mathbb{R}^n, \, q \in \mathbb{R}^1, \, p \in \mathbb{R}^1,
\label{BTham_2}
\end{equation}

\medskip
\noindent
{\bf Step 3: Construction of phase space structures.}
\medskip

We consider  the $2m - 1$ dimensional energy surface $H(\theta, I, q, p) =h$. 
The motivation for rotating the saddle in Eq.\ 
\ref{eq:saddlerot} was to clarify the meaning of reaction -- it corresponds to 
a change in  
sign of the $q$ coordinate.  
Therefore, neglecting the ${\cal O}(3)$ terms, setting $q=0$, the forward and 
backward DSs are given by:
\begin{subequations}
\begin{align}
 p_{\text{f}} &= +\sqrt{\frac{2 }{\lambda (\theta)}}\sqrt{h-\langle \omega, 
 I \rangle - \frac{1}{2} \langle AI, I \rangle} 
 \label{forwardDS} \\
 p_{\text{b}} & = -\sqrt{\frac{2 }{\lambda (\theta)}}\sqrt{h-\langle \omega, 
 I \rangle - \frac{1}{2} \langle AI, I \rangle},
 \label{backwardDS}
 \end{align}
 \end{subequations}
 respectively, and these two DSs meet at the NHIM, $q=0, \, p=0$: 
\begin{equation}
 h=\langle \omega, I \rangle + \frac{1}{2} \langle AI, I \rangle
 \label{NHIM}
 \end{equation}

 These are interesting equations. They show that the DSs vary with $\theta$, but 
 that the NHIM is constant in $\theta$, as follows from 
 the form of Hamilton's equations in these coordinates. 
 At the origin in the space of the saddle degrees of freedom, 
 we have an integrable Hamiltonian system 
 in action-angle variables. We do not attempt to make 
 the hyperbolic DoFs constant by carrying out a Floquet transformation. 
 As a result, the DS (which incorporates hyperbolic DoFs) varies with $\theta$. 
 In some sense, one could view the procedure that gives these coordinates as a 
 type of ``partial normalization'' where we normalize 
 the elliptic DoFs, and leave the hyperbolic DoFs alone.

In this section we have shown that near a partially hyperbolic 
invariant torus a NHIM exists from which a codimension one DS  
can be constructed having the no-recrossing property \cite{waalkens2004direct}. 
We have not provided a method for explicitly constructing the 
coordinates in which we constructed the NHIM and DS. However, 
in the next section we will describe a method for sampling the DS  
for a 2 DoF system and a special 3 DoF system which is relevant 
to the roaming scenarios that we study.

\newpage

\section{A 3 DoF System: A 2 DoF subsystem weakly coupled 
to an elliptic DoF Applicable to Capture Theory in Atom-Diatom Reactions}
\label{sec3}

We consider the  case of a 2 DoF subsystem weakly coupled to an elliptic DoF (i.e., 
a 1 DoF nonlinear oscillator). This situation arises, for example, in 
triatomic molecules when one of the atoms is far from the diatomic fragment, 
and serves as a model for capture
of an atom by a diatomic molecule. In Jacobi coordinates, the DoF
describing the diatomic molecule vibration is, in this situation, 
almost decoupled from the DoF describing the distance 
of the third atom to the center of mass of the diatomic 
and the angle between the diatomic and the third atom.
In the following, we will first specify this model 
more precisely and show how a NHIM can exist in the model. 
We then describe  a numerical procedure
to sample a DS defined by the NHIM.

\subsection{The Existence of a NHIM for a 2 DoF subsystem weakly coupled to an elliptic oscillator}
\label{sec3subsec1}

In order to show the existence of a NHIM we first give a precise description 
of the model to be analyzed. We consider a Hamiltonian describing 
a triatomic molecule in Jacobi coordinates for total angular 
momentum $J=0$ (rotating in the plane) of the following form:
\begin{equation}
H (r,R, \theta,p_r,p_R,p_{\theta}) = \frac{p_r^2}{2\mu_r} + \frac{p_R^2}{2\mu_R} + \frac{p_{\theta}^2}{2} \left( \frac{1}{\mu_R R^2} + 
\frac{1}{\mu_r r^2}  \right)+ V(r,R,\theta),
\label{triatomicHam}
\end{equation}
where $r$ is the diatomic internuclear distance, $R$ is the distance from the 
centre of mass of the diatomic to the third atom and
$\theta$ is the angle between $r$ and $R$, and $p_r$, $p_R$ and $p_{\theta}$ 
are the corresponding conjugate momenta. The assumption that the vibration 
of the diatomic molecule is approximately decoupled from
the $(R,p_R)$ and $(\theta,p_{\theta})$ DoF can be expressed by saying 
that for large atom- diatom distances (small parameter $r/R$)
the potential $V(r,R,\theta)$ has the form:
\begin{equation}
V(r, R, \theta) = V_r (r) + V_R (R) + V_{\text{coupling}} (r,  R,  \theta).
\label{decouppot}
\end{equation}

Using a multipole expansion \cite{Stone2013}, individual terms in the above  equation 
can be written explicitly as:
\begin{subequations}
\begin{align}
V_r(r) & = D_e (1 - \exp[- \kappa (r - r_e)])^2,
\label{defVr} \\
V_R(R) & = -\frac{\alpha_\nu}{R^\nu},
\label{defVR} \\
V_{\text{coupling}} (r,  R,  \theta) & = - \frac{1}{R^\nu} \sum_{k=1,2,\ldots} \, 
\alpha_{\nu + k} \left(\frac{r}{R}\right)^k P_k(\cos \theta) .
\label{defVcoupling}
\end{align}
\end{subequations}
Eq.~(\ref{defVr}) is a Morse potential 
representing the interaction between the two atoms forming the diatomic.
The exponent $\nu =4$ in Eq.~(\ref{defVR}) corresponds to a charge/induced-dipole interaction.
The coupling term Eq.~(\ref{defVcoupling}) is a power series in the ratio $r/R$, 
which is assumed to be small (large $R$).

To show the existence of a NHIM, we first consider the case $V_{\text{coupling}} (r, R, \theta) =0$.
Hamilton's equations in the uncoupled case have the form:
\begin{subequations}
\begin{align}
\dot{r}  &=  \frac{\partial H}{\partial p_r} = \frac{p_r}{\mu_r},  \label{Hameqa} \\
\dot{R}  &=  \frac{\partial H}{\partial p_R} = \frac{p_R}{\mu_R}, \label{Hameqb} \\
\dot{\theta}  &=  \frac{\partial H}{\partial p_\theta}  =  \left( \frac{1}{\mu_r r^2} + \frac{1}{\mu R^2} \right) p_\theta, \label{Hameqc} \\
\dot{p}_r  &=  - \frac{\partial H}{\partial r} = \frac{1}{\mu_r r^3} p_\theta^2 - \frac{\partial V_r}{\partial r} (r), \label{Hameqd} \\
\dot{p}_R  &=  - \frac{\partial H}{\partial R} = \frac{1}{\mu R^3} p_\theta^2 - \frac{\partial V_R}{\partial R} (R), \label{Hameqe} \\
\dot{p}_\theta  &=   -\frac{\partial H}{\partial \theta}  =  0. \label{Hameqf} 
\end{align}
\end{subequations}
Eq.~(\ref{Hameqf}) implies that $p_\theta$ is a constant of the motion. With $p_\theta$ constant, Eqs.~(\ref{Hameqa}), (\ref{Hameqd}), (\ref{Hameqb}) and (\ref{Hameqe}) represent two uncoupled 1 DoF systems for which the solutions ($r(t)$, $p_r(t)$) and ($R(t)$, $p_R(t)$)
can be found separately. Inserting  these solutions into Eq.~(\ref{Hameqc}), the solution $\theta(t)$ can then be determined. The equations of
motion for the uncoupled case (\textit{i.e.} $V_{\text{coupling}} (r, R, \theta) =0$) therefore have the form of 3 uncoupled 1 DoF systems. The DoF ($\theta$, $p_\theta$)
is  an oscillator in the form of an angle-action pair. 

We assume that in the 2 DoF subsystem corresponding to the coordinates 
$( R, \theta,  p_R, p_\theta)$ we have located  an unstable periodic orbit. 
This unstable periodic orbit is the cartesian product of 
a hyperbolic fixed point in the $(R, p_R)$ coordinates, 
denoted $(\bar{R}, \bar{p}_R)$  (the ``centrifugal barrier'' in the 1 DoF $(R, p_R)$ subsystem)
and a periodic orbit in the 1 DoF oscillator 
$(\theta, p_\theta)$ subsystem
which we denote by  $(\theta (t), p_\theta = \rm{constant})$. 
From the previous section, this periodic orbit  is a NHIM for this 2 DoF subsystem.

The NHIM for the 3 DoF uncoupled system for a fixed energy $E$ is obtained 
by taking the cartesian product of the unstable periodic orbit in the 2 DoF 
subsystem with a periodic orbit  in the ($r$, $p_r$) subsystem, 
which we denote by ($\bar{r}(t)$, $\bar{p}_r (t)$). More precisely,
\begin{equation}
\begin{aligned}
\text{NIHM}_{\text{uncoupled}}  =  \{ & (r, R,\theta, p_r, p_R, p_\theta) \, | \, 
\left(r=\bar{r}(t), R=\bar{R}, \theta (t),  p_r=\bar{p}_r (t), p_R=\bar{p}_R , p_\theta \right)=E, \\ 
 & \theta \in S^1, \, p_\theta \in B \subset \mathbb{R} \},
\end{aligned}
\end{equation}
where $B \subset \mathbb{R}$ denotes the  domain of $p_\theta$. 
The NHIM is 3-dimensional in the 5-dimensional energy surface. 
More details of the structure of the NHIM are given in the next section.

We can now appeal to the general persistence theory for NHIMs under perturbation 
\cite{Wiggins94} to conclude that the NHIM just constructed
persists when the term $V_{\text{coupling}} (r, R, \theta)$ is non-zero. 
This completes the proof of the existence of a NHIM for a diatomic molecule 
weakly interacting with an atom for total angular momentum $J=0$.

\subsection{Approximate numerical procedure for sampling the DS}
\label{sec3subsec2}

In the preceding subsection we demonstrated the existence of a NHIM under certain assumptions for a diatomic
molecule weakly interacting with an atom. The assumptions used to demonstrate the 
existence of the NHIM are not
sufficient to provide  a straightforward 
procedure to sample the DS attached to this NHIM. 
We now make an
even stronger assumption that will enable us 
to sample the DS.

Specifically, we now assume that  
for sufficiently large values of $R$ the diatomic vibration is completely decoupled
from the  other DoF. This means that we assume that (for large $R$) 
the coupling term $V_{\text{coupling}} (r, R, \theta)$ in the 
potential function Eq.~\ref{decouppot} will depend only on $R$ and $\theta$, \textit{i.e.} 
$V_{\text{coupling}} (r, R, \theta) \rightarrow V_{\text{coupling}} (R, \theta)$. 
In addition, in the 2 DoF subsystem Hamiltonian we consider the coordinate $r$ to be fixed at its equilibrium
value $r_e$, \textit{i.e.} at the minimum of the Morse like potential well. 
Under these stronger assumptions our system becomes  
a completely decoupled 2 DoF subsystem
plus an elliptic oscillator. 
The sampling of the DS then reduces to the sampling of the part 
of the DS for the 2 DoF subsystem and the sampling of the other
part corresponding to the elliptic oscillator.

In the following we first describe an algorithm for 
sampling a DS attached to a NHIM in a 2 DoF system. 
Then we use this
algorithm to present a procedure for sampling a DS for 
a system composed of the cartesian product of a 2 DoF subsystem
with an elliptic oscillator.

\subsubsection{Sampling of a Dividing Surface attached to a NHIM for a 2 DoF System for a  Fixed Total Energy}
\label{sec3subsec2susubsec1}

When the theory developed in Sec. \ref{sec2} is applied to a 2 DoF system, the hyperbolic torus reduces
to an unstable periodic orbit. In other words, the NHIM in this case is just an unstable periodic orbit. In this section we
describe an algorithm to sample points on a DS attached to a periodic orbit for a 2 DoF system.

For a 2 DoF system the DS is topologically homeomorphic either to a 2-sphere of which the NHIM 
(the PO) is an equator or to a 2-torus (see below).
To sample points on the DS we need to specify the values of four phase space coordinates. 
The PO is generally obtained by a numerical search \cite{Farantos98}. 
The NHIM is represented by discrete points in phase space on  the PO at different
times corresponding to the discretization of the period of the PO. Generally, 
once projected onto configuration space, these points lie on a curve. This curve in
configuration space can be approximated by a spline interpolation in order to 
obtain a continuous representation of the curve. 
The sampling of the DS starts by sampling points along this curve in configuration space. 
This fixes the values of the
2 configuration space coordinates.  To determine the values of the remaining 2 phase space coordinates 
(momenta), we use the fact that 
 the sampling is at a fixed total energy. 
 So, for each
point on the projection of the PO into configuration space we can 
scan the value of one of the momenta 
and then obtain the remaining momentum by solving numerically 
the equation $H=E$, $H$ being the Hamiltonian of the 2 DoF system, 
for the remaining momentum coordinate. The
maximum value of the first momentum coordinate to be 
scanned can be determined by setting the second momentum
coordinate to zero and solving the equation $H=E$ for each point of the PO in configuration space.

To specify such an algorithm to sample points on the DS, 
we denote configuration space coordinates by $(q_1,q_2)$ 
and momentum space coordinates by $(p_1,p_2)$. (This notation emphasizes 
that our approach is applicable to systems other than the atom-diatom model considered
here.) The algorithm to sample the DS for a 2 DoF system can be summarized as follows:
\begin{enumerate}

\item Locate an unstable PO.

\item Perform  a spline interpolation \cite{Press92} of the curve obtained by projecting 
the PO into configuration space.

\item Sample points on that curve and obtain points $(q_{1i},q_{2i})$, for $i=1$ to 
the number of desired points.

\item For each point $(q_{1i},q_{2i})$ determine $p_{1max}$ by solving 
$H(q_{1i},q_{2i},p_1,0)=E$ for $p_1$.

\item For each point $(q_{1i},q_{2i})$ sample $p_1$ from zero to $p_{1max}$ 
and solve the equation $H(q_{1i},q_{2i},p_{1j},p_2)=E$ to
obtain $p_2$.

\end{enumerate}

For kinetic energies that are quadratic in the momenta, 
the momentum values satisfying the energy condition $H=E$ above will appear with plus or minus
signs. Computing the points of the DS with both  positive 
and negative for $p_1$ covers both hemispheres of the DS.

For a DS is obtained by this sampling procedure, we now address the following questions:
\begin{itemize}

\item What is the dimensionality of the sampled surface?

\item What is the topology of the sampled surface?

\item Does the NHIM (PO) bound the 2 hemispheres of the DS?

\item Do the trajectories cross the DS and what does `reaction' mean with respect to  the sampled DS?

\end{itemize}

Consider the Hamiltonian
\begin{subequations}
\begin{align}
H(q_1,q_2,p_1,p_2) & = T(q_1,p_1,p_2)+V(q_1,q_2) \\
& =
\frac{p_1^2}{2\mu}+\frac{p_2^2}{2}\left(\frac{1}{I}+\frac{1}{\mu q_1^2}\right)+V(q_1,q_2).
\end{align}
\end{subequations}
Moreover, the potential is periodic in coordinate $q_2$ with period $2\pi$.

As we have noted, the sampling procedure begins with the location of 
an unstable PO, where the projection of 
the located PO is 1-dimensional.

Now we turn our attention to the questions listed above.

\medskip
\paragraph{Dimensionality}

The dimensionality of the sampled DS is relatively straightforward to determine.
As described above, we locate an unstable PO and perform a 
spline interpolation for the set of points obtained
by projecting the PO into configuration space. We then sample points along this
curve. For each $(q_1,q_2)$ point we determine the maximum possible value
of one of the momenta by solving the equation $H(q_1,q_2,p_1,0)=E$ for $p_1$,
for example. We obtain the maximum value for $p_1$, $p_{1max}$, 
permitted by the energy constraint. 
We then sample the momentum $p_1$  from $-p_{1max}$ to $+p_{1max}$.

Topologically, the space sampled in configuration space is a 1D segment and the space sampled
in the momentum is also a 1D segment. The full space sampled is therefore  
a 2D surface embedded in the 3D energy surface in the 4D phase space.

\medskip
\paragraph{Topology of the DS}

We must distinguish 2 cases, which differ in the type of PO sampled.
There are 2 types of POs we must consider. The first type is a PO for which
the range of the PO in the periodic coordinate $q_2$ is less than the full period 
of the coordinate. The second type is a PO which makes a full cycle in the periodic coordinate.

\medskip
\noindent
\subparagraph{Type 1 PO}

As we noted, this type of PO has a  range in the periodic coordinate  
that is less than the period of the coordinate. These POs have 
2 turning points at which the total energy is purely potential 
and the momenta are zero.

If, for example, we sample the $q_2$ coordinate, this also fixes $q_1$. 
On a fixed energy surface we have to satisfy the equation 
$H(q_1,q_2,p_1,p_2)=E$, so having fixed $q_1$ and $q_2$ the potential energy is fixed
at a particular value $V_0$. The equation to be satisfied by the momenta  is then
\begin{equation}
\frac{p_1^2}{2\mu}+\frac{p_2^2}{2}\left(\frac{1}{I}+\frac{1}{\mu q_{10}^2}\right)+V_0=E .
\end{equation}
This is the equation of an ellipse in the  $(p_1,p_2)$ plane. 
Moreover, the lengths of the large and small
axes of the ellipse tend to zero as we approach the turning points where $E=V_0$. 
So, as we follow the curve of the PO in configuration space from one turning point to the other, 
the intersection of the DS with the $(p_1,p_2)$ plane begins as a point  
and then becomes a family of 
ellipses whose 2 axes depend
on the potential energy value, eventually shrinking down to a point again as  
the other turning point is reached. The topology
of the DS is then the product of the 1D segment in configuration space with  a 
family of ellipses degenerating to points at either end: that is, a 2-sphere.

\medskip
\noindent
\subparagraph{Type 2 PO}

This  type of PO  makes a full cycle in the periodic coordinate. 
It is important to note that for this case the DS is constructed from 2 POs. 
For a PO that makes
a full cycle in the periodic coordinate $q_2$, the conjugate momentum $p_2$ never vanishes 
and always has the same sign (either positive or negative). 
For a given PO there is a twin PO for which the
momentum has the opposite sign (the PO trajectory  
traverses the same path but in the opposite direction).

When we sample the curve in configuration space and analyse the shape of the DS in momentum
space for a particular point in configuration space, 
we solve the same equation as for the previous 
case. Therefore, the shape of the DS in momentum space is also an ellipse. 
However, now the ellipses never degenerate to points
as there are no turning points along the PO at which the total energy 
is purely potential. Also, since the coordinate $q_2$ is periodic, the end
point of the PO has to be 
identified with the starting point.
In this situation the DS is then the cartesian product 
of a circle with a family of ellipses and has the topology
of a 2-torus.

\medskip
\paragraph{The NHIM separates the DS into two parts}

To discuss this point we also must distinguish between type 1 and type 2 POs.

\medskip
\noindent
\subparagraph{Type 1 PO}

As above we fix a particular point in configuration space $(q_{10},q_{20})$. 
As we scan one of the
momenta there will be a particular value corresponding to the
momentum value  on the PO. Since, the sampling is at the same energy as the PO energy,
the value of the remaining momentum  will also correspond to the value 
on the PO. For the configuration space point $(q_{10},q_{20})$ this happens exactly
twice. For a nearby configuration space point $(q_{10}',q_{20}')$ the 2 momentum space points
belonging to the PO will be near the 2 points for the point $(q_{10},q_{20})$. 

As we move along the PO
in configuration space these 2 points in momentum space will trace out the PO 
in phase space. 
At the turning
points, the ellipse in momentum space shrinks to a point which is
also the meeting point of the 2 points
in momentum space belonging to the PO. So, the NHIM (the PO) belongs to the 2D sphere we sample.
It forms a 1D closed curve which divides the
sphere into 2 hemispheres.

\medskip
\noindent
\subparagraph{Type 2 PO}

For this type of PO, if we fix a configuration space point $(q_{10},q_{20})$ and scan one of the
momentum variables, we will meet a point for which the momentum equals the value of
that momentum on the PO. The other momentum has its value equal 
as its value on the PO due to the energy constraint. We will 
also encounter a point for which the value of the momentum we are scanning
is equal to the value of that momentum coordinate on the twin PO (the one which moves in the  opposite
direction with $p_2$ having the opposite sign of the original PO). 
The corresponding value of the remaining
momentum coordinate is also on the twin PO.

\begin{figure}
\includegraphics[scale=0.4]{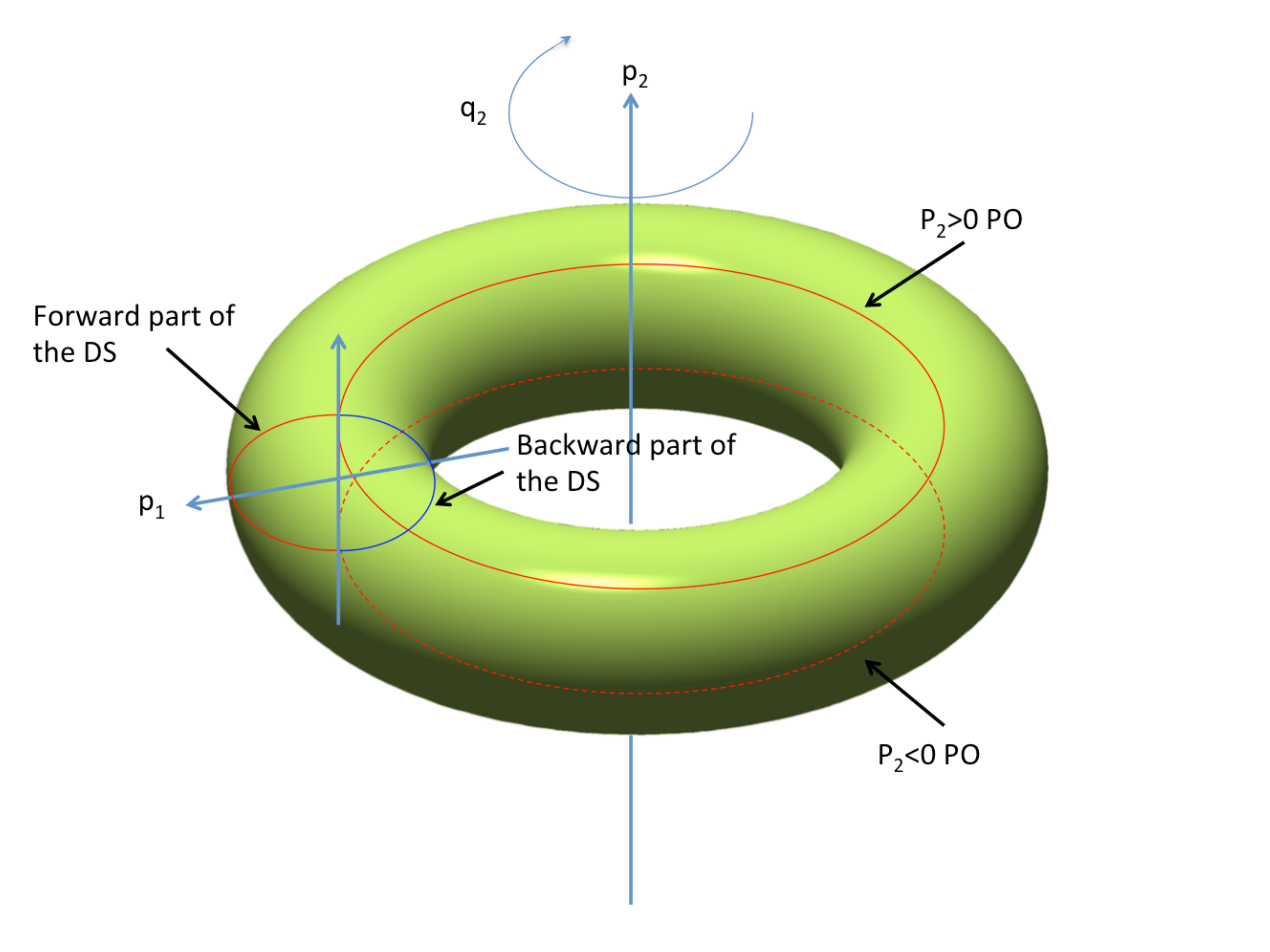}
\caption{\label{fig1} Topology of the DS for type 2 POs. }
\end{figure}

The 2 POs belong to the surface we are sampling. Taken individually, these two curves do not divide
the 2-torus into two parts but the DS is divided into two parts when we considering 
both POs together: one part of the DS
connects the $p_2>0$ PO with the $p_2<0$ PO on the left of the ellipse in momentum space
and one part connects the $p_2>0$ PO with the $p_2<0$ PO on the right of the ellipse, 
see Fig.~\ref{fig1}.

\medskip
\paragraph{Trajectories cross the DS and the meaning of reaction for the constructed DS}
\smallskip

Except for points on the NHIM, which belongs to the surface we are sampling, 
no trajectory initiated on the DS
remains on the DS. Off the NHIM, the Hamiltonian
vector field is everywhere transverse to the sampled surface, and 
every trajectory initiated on the DS has to leave the surface, i.e., 
trajectories cross the DS.

We have shown that the NHIM (or the 2 NHIMs for type 2 POs) 
divides the DS into two parts. One
part of the DS intersects trajectories evolving from reactants to products 
and the other intersects trajectories
evolving from products to reactants. Here, `reaction' means 
crossing one of the two parts of the DS and the
direction of the reaction, forward or backward, 
depends on which half of the DS the trajectory crosses.

To show that the vector field is nowhere tangent to the sampled surface, 
except on the NHIM, we must evaluate the flux form \cite{Ezra09} 
associated with the Hamiltonian vector field through the DS on
tangent vectors to the DS. The Hamiltonian vector field is not tangent to the DS 
if the flux form is nowhere zero on the DS, except at the NHIM.
Details of this calculation are given in Appendix \ref{app:transverse}.

\subsubsection{Sampling of a Dividing Surface Attached to a NHIM for a 2 DoF Subsystem Plus an Elliptic Oscillator at a Fixed Total Energy}
\label{sec3subsec2subsubsec2}

Under the assumptions introduced at the beginning of this section, 
the problem of the capture of an atom by a diatomic
molecule at zero total angular momentum reduces to consideration of the dynamics of 
a 2 DoF subsystem and 
an elliptic oscillator.
The algorithm presented for the sampling of a DS attached to a PO  
will be an essential element in our derivation of a sampling procedure of the DS in the case
of a 3 DoF system composed of a 2 DoF subsystem plus an elliptic oscillator.

To begin, we need to understand the nature of the NHIM for this case. 
The motion for the elliptic oscillator (the diatomic vibration) consists of
periodic orbits. For sufficiently small amplitude vibration (energy below the 
dissociation energy of the Morse like potential), these
POs are stable. The NHIM for the full 3 DoF system (2 DoF subsystem plus 1 DoF elliptic oscillator) 
is obtained by considering the direct
product of the NHIM for the 2 DoF subsytem with a PO of the elliptic oscillator. 
The NHIM for the 2 DoF subsytem, as we saw above, is just
an unstable PO. If we consider a particular unstable PO for the 2 DoF subsystem 
and a particular PO for the elliptic oscillator, the structure
obtained by the cartesian product of these 2 manifolds is a 2-dimensional torus. 
Every particular partitioning of the total energy $E$ between the 2 DoF subsystem and
the elliptic oscillator corresponds to a particular 2-dimensional
torus for the NHIM. The NHIM is then made up of a 1-parameter family of 2-dimensional tori, 
where the parameter of this family determines the 
distribution of energy between the 2 DoF subsystem and the elliptic oscillator.

Now that we have the structure of the NHIM it remains
to construct the associated DS. Like the NHIM, the
DS has a product structure: one part is associated with the 2 DoF subsystem 
and the other with the elliptic oscillator. For the elliptic oscillator
all phase space points belong to a particular PO. For the 2 DoF
subsystem the component of the DS consists of a DS attached to a PO 
as we described in Sec. \ref{sec3subsec2}. 
If we consider one particular partitioning of the total energy $E$ 
between the 2 DoF subsystem and the elliptic oscillator, the portion of the 
DS associated with this distribution
consists of the direct product of the DS associated with the unstable PO for the 2 DoF subsystem 
and the PO for the elliptic oscillator. 
The full DS is obtained
by considering the DS for all possible partitions of the energy. 
The full DS is then foliated by `leaves', where each 
leaf of the DS is associated with a particular partitioning of the energy.

To implement an algorithm for sampling the DS we need to determine
the potential $V_r(r)$ for the diatomic molecule. 
This can be done by
fixing a sufficiently large value of $R$, and fixing a value for $\theta$. In general, for large values of $R$
the potential $V(r,R,\theta)$ is isotropic, so any value of $\theta$ is sufficient. 
One can
then use spline interpolation to obtain an analytical representation of the 
resulting Morse-like potential. 

An algorithm to sample the DS in the case of a decoupled 2 DoF subsystem plus an
elliptic oscillator is given as follows:

\begin{enumerate}

\item Consider a total energy $E$ for the full system.

\item Consider a particular partitioning of the total energy $E$ 
between the 2 DoF subsystem and the elliptic oscillator.

\item Sample points on the PO corresponding to the energy associated with  the elliptic oscillator.

\item Sample points on the DS attached to the PO corresponding to the energy 
associated with the 2 DoF subsystem using the algorithm
presented in the Sec. \ref{sec3subsec2}.

\item Take the direct product the two sets of points obtained in the two previous steps.

\item Repeat the previous steps for many different energy distributions 
between the 2 DoF subsystem and the elliptic oscillator.

\end{enumerate}

There is a subtlety related to the sampling procedure just described. 
The hyperbolic POs of the 2 DoF subsystem are 
associated with the centrifugal barrier arising in the atom-diatom
Hamiltonian in Jacobi coordinates. 
Following the family of these POs it is found that 
as the energy decreases their location in the  $(R,\theta)$ plane moves to 
larger and larger values of $R$.
This suggests that the family goes asymptotically to a  
neutrally stable PO located at $R \to \infty$ \cite{neutrally_stable_footnote1}. 
Reducing  the energy in the 2 DoF subsystem therefore means locating
POs at larger and larger distances in $R$, with longer and longer periods. 
Location of all
such POs is clearly not possible numerically so
that the sampling procedure must stop at some finite (small) energy
in the diatomic vibration. 
The sampling procedure will nevertheless cover the DS, at least partially.

Alternatively, the energy  in the diatomic vibration can simply be fixed
at a specific value, as in the standard quasiclassical 
trajectory method (see, for example, the work of Bonnet et al. 
\cite{Bonnet1997183} in which the energy of the diatomic 
vibration is fixed at the zero point energy).
This approach then samples a particular leaf of the classical DS.

\newpage

\section{An Example: Recombination of the Ozone Molecule}
\label{sec4}

The recombination reaction O $+$ O$_{2}$ $\rightarrow$ O$_{3}$ 
in the ozone molecule
is known to exhibit an unconventional isotope effect \cite{janssen2001kinetic}. 
There is a large literature on this
subject and many theoretical models have been proposed to identify 
the origin of this isotope effect \cite{gao2001strange,schinke2006dynamical,janssen2001kinetic}. 
One key to understanding the isotope effect in ozone is to
model the recombination of an oxygen molecule with an oxygen atom. 
Marcus and coworkers have made a detailed study of the isotope
effect in ozone \cite{gao2001strange,gao2002theory,gao2002theoretical} 
and have used different models for the transition
state in the recombination of ozone (tight and loose TS).
In their studies, they used a variational version of RRKM theory \cite{marcus1952unimolecular,Truhlar_Garrett1984}
(VTST) to locate the transition state. 
In order to obtain agreement between their calculations and
experimental results they introduced parameters whose role consisted of 
correcting the statistical assumptions inherent in RRKM theory. 
Marcus and coworkers proposed several 
explanations for the physical origin of these parameters 
\cite{gao2001strange,gao2002theory,gao2002theoretical} 
but their significance is still not fully understood. 
The deviation from statistical beheviour in the ozone molecule
suggests that a dynamical study of the recombination reaction of ozone 
rather than a statistical model 
might shed light on the origin of the 
unconventional isotope effect. Such
a study necessitates 
the use of DSs having dynamical significance, i.e., constructed in phase space.

To illustrate the construction of phase space DS for the 
capture of an atom by a diatomic molecule we
apply the algorithms described in the previous section 
to the ozone recombination reaction. To attack this problem
we require a PES for the ozone molecule which describes large amplitude motion. 
Schinke and coworkers produced the first
accurate PES for ozone\cite{siebert2001spectroscopy,siebert2002vibrational}. 
This PES exhibits a small
barrier in the dissociation channel just below the asymptote for dissociation 
forming a van der Waals minimum and
an associated `reef structure'. 
Recent \textit{ab initio} calculations \cite{Tyuterev2013,Dawes2011,Dawes2013} 
have however shown 
that the barrier is not present,  and that the potential
increases monotonically along the reaction coordinate 
(there is no saddle point along the reaction coordinate).
Here, we use the PES produced by Tyuterev and coworkers\cite{Tyuterev2013}, 
which has recently been used to
interpret experimental results \cite{Tyuterev2014}.

In the following we treat the recombination of ozone molecule for the isotopic combination 
$^{16}$O$^{16}$O$^{16}$O by using two models. 
First, we treat the problem with a reduced 2 DoF 
model with zero total angular momentum, and then we look at the 
3 DoF problem  also with  zero angular momentum.

\subsection{2 DoF model}
\label{sec4subsec1}

The 2 DoF case is easily treated using the analysis of Sec. \ref{sec3subsec1}
by freezing the diatomic vibration, i.e., ($r, p_r$) variables.
We therefore fix the diatomic bond length at its equilibrium 
value, $r_e$, and set $p_r=0$. 
The dependence on $r$ in the coupling term of the potential then drops out 
and we have
simply $V_{\text{coupling}} (R,  \theta) = V_{\text{coupling}} (r=r_e,  R,  \theta)$. 
The NHIM is obtained as previously by considering
the uncoupled case with $V_{\text{coupling}} (R,  \theta)=0$ 
and a hyperbolic equilibrium point in the ($R, p_R$) DoF,
and is just a periodic orbit 
in the ($\theta, p_\theta$) DoF. This NHIM persists 
for non-zero coupling and consists of a periodic orbit 
where variables ($R, p_R$) and ($\theta, p_\theta$) are in general coupled.

The sampling of the DS follows the algorithm presented in Sec. \ref{sec3subsec2susubsec1}. 
We use the sampled points on the DS
to initiate trajectories and follow the
approach of the atom to the diatomic molecule. The resulting DS is
a 2-dimensional surface in phase space whose points belong to the 
chosen energy surface. For our
simulation we choose the total energy of 9200 cm$^{-1}$. 
The DS constructed
from our algorithm is the first `portal'  the system has to cross in order 
for the atom to interact with the diatomic molecule. 
This DS is then 
the phase space realisation of the so-called Orbital or loose Transition State (OTS) \cite{Wiesenfeld03,Wiesenfeld05}. 
At the so-called Tight Transition State (TTS) \cite{Mauguiere_2014cpl},
the coordinate $\theta$ undergoes constrained small oscillations. 
In our phase space setting, these TTS for a 2 DoF system are DSs attached to unstable POs. 
In the following the terms OTS and TTS are to be understood as synonymous with 
DSs constructed from NHIMs (POs for a 2 DoF system). 
An important mechanistic issue arising 
in modeling a reaction is to determine whether the reaction is 
mediated by an OTS or a TTS. However, in many situations both types of TS  
co-exist for the same energy and one has to understand 
the dynamical implications of the presence of these 2 types of TS. 

In a recent work we analysed this question for a model ion-molecule system  
originally considered by Chesnavich \cite{Chesnavich1981,Chesnavich82,Chesnavich1986}.
We established a 
connection between the dynamics induced by the presence of these 2 
types of TS and the phenomenon of roaming \cite{Mauguiere_2014cpl,Mauguiere_2014jcp}. 
Specifically,  our phase space approach enabled us to define unambiguously 
a region of phase space bounded by the OTS and TTS
which we called the \emph{roaming region}. We were also able to classify 
trajectories initiated on the OTS into four different classes. These four classes were derived
from two main categories. The main category consists of
reactive trajectories. These trajectories start at the OTS and enter the roaming region
between the OTS and TTS. They can then exhibit
complicated dynamics in the roaming region and then react 
by crossing the TTS to form a bound triatomic molecule.
Within this category we distinguish between \emph{roaming} 
trajectories 
making some oscillations in the roaming region (oscillations were defined by the presence
of turning points in $R$ coordinate) and 
\emph{direct} trajectories that do not oscillate 
in the roaming region. 
The other main category of trajectories consists of trajectories that do not react 
and for which the atom finally separates
from the diatomic molecule. 
Again we distinguish between trajectories which ``roam" 
(making oscillations in the roaming region) and 
those that escape directly by bouncing off 
the hard wall at small values of $R$. 

\begin{figure}
\includegraphics[scale=0.7]{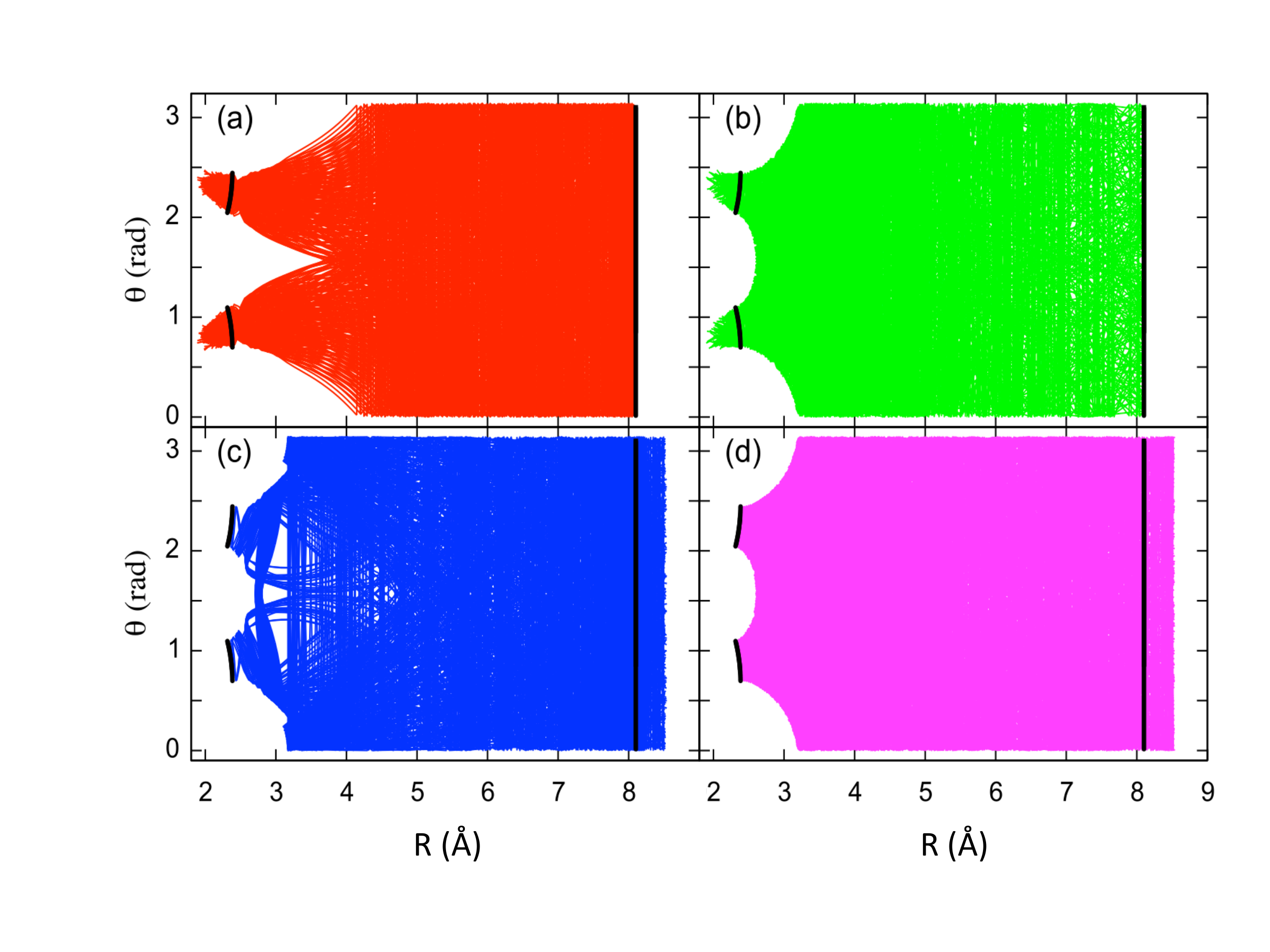}
\caption{\label{fig2} Trajectories propagated 
from the OTS for the 2 DoF model projected on configuration space.
The black line at $R \simeq 8 \text{\AA}$  is the PO from 
which the OTS is constructed. The short black curves
at  $R \simeq 2.5 \text{\AA}$  are the POs from which the 
TTS are constructed. (a) Direct reactive trajectories.
(b) Roaming reactive trajectories. (c) Direct non reactive trajectories. 
(d) Roaming non reactive trajectories.}
\end{figure}

For the ozone recombination problem 
reduced to a 2 DoF model we employ the same classification. 
Figure ~\ref{fig2} shows the four classes of trajectories obtained by
propagating incoming trajectories originated at 
the OTS. The black line at $R \simeq 8 \text{\AA}$ 
is the PO from which the OTS is constructed. The two black curves
at $R \simeq 2.5 \text{\AA}$ are the two POs from which the two TTS are constructed. 
These two TTS correspond to the two channels by which the
ozone molecule can recombine. For the case of $^{16}$O$_3$ the 
two channels reform the same ozone molecule but for different isotope composition
these channels may lead to different ozone molecules. 
These trajectories show  that the roaming phenomenon is at play in the 
ozone recombination problem. It would be very interesting to investigate 
how the roaming dynamics varies as the masses of the different oxygen atoms change,
as this may provide new insight into the isotope effect in ozone recombination.

\subsection{3 DoF Model}
\label{sec4subsec2}

We now consider the problem of ozone recombination 
in a 3 DoF model with zero total angular momentum. 
Addition of angular momentum DoF is in principle possible 
in our phase space description of reaction dynamics, 
but consideration of these DoF will increase the
dimensionality of the problem 
and will require a more elaborate sampling procedure. 
Consideration
of angular momentum in capture problems have been investigated
by Wiesenfeld and coworkers\cite{Wiesenfeld03,
Wiesenfeld04a,Wiesenfeld05} and recently by MacKay and Strub\cite{MacKay2015}.

The construction of the DS for the OTS follows the procedure 
described in Sec. \ref{sec3subsec2subsubsec2}. As
explained in Sec. \ref{sec3subsec2subsubsec2} we cannot sample the full DS. 
In our simulation we sample one 
leaf of the DS by considering a single partitioning of the energy 
between the two DoF subsystem and the diatomic vibration.
Here, for the sampling of the DS we fixed the diatomic length $r$ to its equilibrium value,
and use the same value of the total energy 
as for the 2 DoF model, 9200 cm$^{-1}$.

\begin{figure}
\includegraphics[scale=0.7]{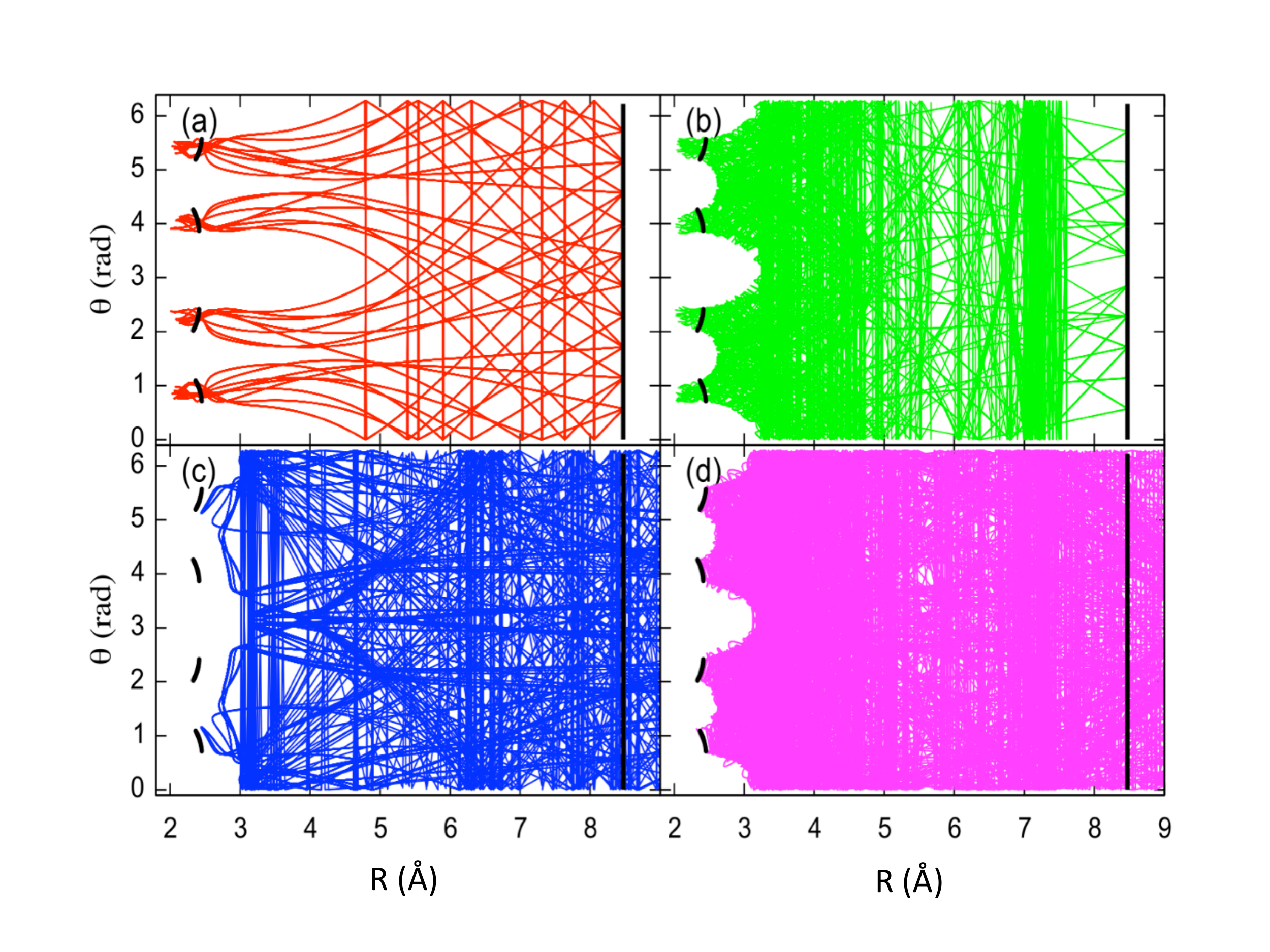}
\caption{\label{fig3} Trajectories propagated from the OTS for the 3 DoF model 
projected on configuration space.
The black line at $R \simeq 8 \text{\AA}$  is the projection of 
the PO from which the OTS is constructed. 
The short black curves
at  $R \simeq 2.5 \text{\AA}$  are projections of POs that may be used as 
approximations of the TTS (see text). (a) Direct reactive trajectories.
(b) Roaming reactive trajectories. (c) Direct non reactive trajectories. 
(d) Roaming non reactive trajectories.}
\end{figure}

The sampled points on the DS were again used to propagate trajectories, which
are classified as previously into four classes.
The result of the propagation is shown on Fig.~\ref{fig3}. 
Again the black line at $R \simeq 8 \,\text{\AA}$ represents the projection 
of the PO defining the OTS, while
the two black curves at $R \simeq 2.5 \,\text{\AA}$ are projections of POs. 
These POs are not strictly 
the proper phase space structures from which TTSs can be constructed as for 
a 3 DoF system the NHIM supporting
a DS consists of a 1-parameter family of invariant 2-tori. 
Nevertheless, these POs give an idea of the approximate 
location of the projection of these structures into 
configuration space  for ozone
recombination. For the ozone problem it appears that 
they are very good approximation to the true NHIMs for the TTSs as we can see
in Fig.~\ref{fig3} that none of the non reactive trajectories (panels (c) and (d)) 
extends  beyond the projections of those POs on configuration
space. The reason for that is presumably that for ozone the diatomic 
vibration is well decoupled from the 2 other DoF throughout
the entire roaming region.
This fact validates the reduced 2 DoF model presented above.

\newpage

\section{Conclusion}
\label{sec:conclusions}

In this paper we have examined
the phase space structures that govern reaction dynamics 
in the absence of critical points
on the PES. We showed that in the vicinity of hyperbolic invariant tori 
it is possible to define phase space dividing surfaces  
that are analogous to the dividing surfaces governing 
transition from reactants to products near a critical point of the PES.

We investigated the problem of capture of an atom by a diatomic molecule 
and showed that a normally hyperbolic invariant manifold 
(NHIM) exists at large
atom-diatom distances, away from any critical points on the
potential. This NHIM is the anchor for the construction of a DS in phase space,
which is the entry `portal' through which the atom has to pass in order to interact 
with the diatomic molecule.  This DS defines 
the outer or loose TS (OTS) governing capture dynamics.

Making certain assumptions, we presented an algorithm 
for sampling an approximate capture DS. 
As an illustration of our methods we applied the algorithm
to the recombination of the ozone molecule. We treated
both 2 and  3 DoF models with zero total angular momentum. 
The co-existence of the OTS and
TTS in ozone recombination means that roaming dynamics 
is observed for this reaction. Such roaming dynamics may
have important 
consequences for the unconventional isotope effect in ozone formation.

\acknowledgments
FM, PC and SW  acknowledge the support of the  Office of Naval Research (Grant No.~N00014-01-1-0769),
the Leverhulme Trust. 
BKC, FM, PC and SW  acknowledge the support of
Engineering and Physical Sciences Research Council (Grant No.~ EP/K000489/1).
GSE and ZCK acknowledge the support of the National Science Foundation under Grant  No.\ CHE-1223754.
We thank Prof. Vladimir Tyuterev for providing us with his PES for ozone.

\appendix

\section{Verification that trajectories are transverse to the DS associated with the NHIM}
\label{app:transverse}

To carry out this calculation we require a parametrization of the DS 
that will enable us derive expressions for tangent vectors to the DS. 
We can then compute the flux form and evaluate it on tangent vectors to the DS.

\medskip
\noindent
{\em Parametrization of the DS}
\smallskip

The DS is a 2-dimensional surface and therefore we require  2 parameters to 
parametrize it. These 2 parameters appear naturally from the sampling method, where
we sample in the periodic coordinate $q_2$ and then in the momentum
coordinate $p_2$. So, we use parameters ($p_2,q_2$). Spline interpolation gives the function
$q_1=q_1(q_2)$ describing the dependence of
coordinate $q_1$ on $q_2$ in the projection of the NHIM onto configuration space. 
The remaining momentum coordinate, $p_1$, is obtained by imposing the constant energy constraint 
$H=E$. The parametrization $\Phi$ of the DS can then be
specified as:
\begin{equation}
\Phi(p_2,q_2)=\left( p_1=\pm \left[ \left( E-V(q_1(q_2),q_2)-\frac{p_2^2}{2I_1(q_1(q_2))} \right) 2\mu \right]^{\frac{1}{2}},
q_1(q_2),p_2,q_2 \right),
\end{equation}
with $1/I_1=1/I+1/(\mu q_1^2)$.

Tangent vectors to the DS are obtained by differentiating the parametrization with respect to the parameters
$p_2$ and $q_2$:
\begin{subequations}
\begin{align}
u_1 & = \frac{\partial \Phi}{\partial p_2} = \left( \frac{\partial p_1}{\partial p_2}, 0, 1, 0 \right)^T =
\left( -\frac{\mu p_2 }{I_1 p_1(p_2,q_2)},0,1,0 \right)^T \\
u_2 & =  \frac{\partial \Phi}{\partial q_2} = \left( \frac{\partial p_1}{\partial q_2},\frac{d q_1}{d q_2} , 0, 1 \right)^T =\left(
 - \frac{ \left( \partial V/\partial q_2 - \frac{p_2^2 (dI_1/dq_2) }{2 I_1^2(q_2)} \right) \mu }
{ p_1(p_2,q_2)},
\frac{d q_1}{d q_2} , 0, 1 \right)^T.
\end{align}
\end{subequations}

\medskip
\noindent
{\em The flux form associated with the Hamiltonian vector field}
\smallskip

For a 2 DoF system the phase space volume form $\Omega$ is expressed using the symplectic
2-form $\omega$ as follows:
\begin{equation}
\Omega=\frac{1}{2}\omega \wedge \omega.
\end{equation}

\noindent
The energy surface volume 3-form $\eta$ is defined by:
 \begin{equation}
\Omega=dH \wedge \eta.
\end{equation}

The Hamiltonian vector field is:
\begin{equation}
X_H=\left( -\frac{\partial H}{\partial q_1}, \frac{\partial H}{\partial p_1}, -\frac{\partial H}{\partial q_2},
\frac{\partial H}{\partial p_2} \right).
\end{equation}
The symplectic 2-form $\omega$ when applied to an arbitrary vector $\xi$ and the vector field
$X_H$ gives:
\begin{equation}
\omega(\xi,X_H)=dH(\xi).
\end{equation}

The flux 2-form $\varphi$ through a codimension one surface in the energy surface (that is a 2-dimensional surface like our DS) is simply given by
the interior product of the energy surface volume form $\eta$ with the Hamiltonian vector field:

\begin{equation}
\varphi=i_{X_H}\eta.
\end{equation}

To compute $\varphi$ we take two vectors $\xi_1$ and $\xi_2$ tangent to the DS and a
vector $\xi_3$ such that $dH(\xi_3)\neq 0$. First we note that the volume form $\Omega$
when applied to $\xi_1$, $\xi_2$, $\xi_3$ and $X_H$ gives:
\begin{subequations}
\label{eqvol1}
\begin{align}
\Omega(\xi_1,\xi_2,\xi_3,X_H) &= \omega(\xi_1,\xi_2) \omega(\xi_3,X_H) -
\omega(\xi_1,\xi_3) \omega(\xi_2,X_H) + \omega(\xi_1,X_H) \omega(\xi_2,\xi_3)  \\
&= \omega(\xi_1,\xi_2) dH(\xi_3) - \omega(\xi_1,\xi_3) dH(\xi_2) + 
\omega(\xi_2,\xi_3) dH(\xi_1) \\
&= \omega(\xi_1,\xi_2) dH(\xi_3).
\end{align}
\end{subequations}
The last line follows since $\xi_1$ and $\xi_2$ are 
tangent vectors to the DS and therefore $dH(\xi_i)=0$ for $i=1,2$. 
Also, we have $dH(X_H)=0$ as $X_H$ is a tangent vector field to the energy surface. 
From the definition of the energy surface volume form we have:
\begin{subequations}
\label{eqvol2}
\begin{align}
\Omega(\xi_1,\xi_2,\xi_3,X_H) &=  dH \wedge \eta (\xi_1,\xi_2,\xi_3,X_H) \\
&=  dH(\xi_1)\eta(\xi_2,\xi_2,X_H) + dH(\xi_2)\eta(\xi_1,\xi_3,X_H) \nonumber \\
& + dH(\xi_3)\eta(\xi_1,\xi_2,X_H)
+ dH(X_H)\eta(\xi_1,\xi_2,\xi_3) \\
& =  dH(\xi_3) \eta(\xi_1,\xi_2,X_H).
\end{align}
\end{subequations}

Since 
$\varphi(\xi_1,\xi_2)=i_{X_H}\eta(\xi_1,\xi_2) = \eta(\xi_1,\xi_2,X_H)$, 
it then follows  from (\ref{eqvol1}) and (\ref{eqvol2}) that:
\begin{equation}
dH(\xi_3) \omega(\xi_1,\xi_2) = dH(\xi_3) \, i_{X_H}\eta(\xi_1,\xi_2) 
\end{equation}
and
\begin{equation}
\varphi = i_{X_H}\eta(\xi_1,\xi_2)=\omega(\xi_1,\xi_2).
\end{equation}
The flux form through the DS is simply the symplectic 2-form applied to tangent vectors
of the DS.

\medskip
\noindent
{\em Condition for non-tangency of the Hamiltonian vector field}
\smallskip

To check that the vector field is not tangent to the DS (\textit{i.e.} trajectories
cross the DS) we need to evaluate the flux form on tangent vectors to the DS.

The calculation we need to carry out is the following:
\begin{subequations}
\begin{align}
\omega(u_1,u_2) &= (dp_1 \wedge dq_1+dp_2 \wedge dq_2) (u_1,u_2)  \\
& =  (u_1^{p_1} u_2^{q_1} - u_1^{q_1} u_2^{p_1}) + (u_1^{p_2} u_2^{q_2} - u_1^{q_2} u_2^{p_2})\\
& =  -\frac{\mu p_2}{I_1 p_1} \frac{dq_1}{dq_2}+1
\end{align}
\end{subequations}
The flux through the DS is then zero whenever:
\begin{equation}
\frac{dq_2}{dq_1} = \frac{p_2/I_1}{p_1/\mu} = \frac{\dot{q_2}}{\dot{q_1}}
\end{equation}
That is, the ratio of velocities corresponds precisely to the
phase point being on the NHIM (PO).


\begin{thebibliography}{80}
\bibitem[{neu()}]{neutrally_stable_footnote1}
\bibinfo{note}{{A periodic orbit is said to be neutrally stable if all of its
  Lyapunov exponents are zero.}}
\end{thebibliography}


\newpage

\end{document}